\documentclass[structabstract]{aa}  
\usepackage{graphicx}
\begin{document}
\title{The binary nature of the  Galactic Centre X-ray source \object{CXOGC J174536.1-285638}.}

\author{J.~S.~Clark\inst{1}
\and P.~A.~Crowther\inst{2}
\and V.~J.~Mikles\inst{3}}
\institute{
$^1$Department of Physics and Astronomy, The Open 
University, Walton Hall, Milton Keynes, MK7 6AA, UK\\
$^2$ Department of Physics \& Astronomy, University of Sheffield, Sheffield, S3 7RH, UK\\
$^3$ Deaprtment of Physics \& Astronomy, Louisiana State University, 273 
Nicholson Hall, Tower Drive, Baton Rouge, LA 70803, USA
}

 \abstract{The combination of X-ray and near-IR surveys of the central 
2$^{\rm o}\times0.8^{\rm o}$ of the Galactic Centre have revealed a population of X-ray bright massive stars. 
However, the nature of the X-ray emission, originating in  wind collision zones or via accretion onto compact objects, 
is uncertain.}{In order to address this we investigated the nature of one of the 
most luminous X-ray sources -  CXOGC J174536.1-285638.}{This was accomplished by an analysis
of the near-IR spectrum with  a non-LTE model atmosphere code to determine the physical parameters of the primary.}
{This was found to be  an highly luminous WN9h star, which is 
 remarkably similar to the most massive stars found 
in the Arches cluster, for which comparison to 
evolutionary tracks suggest an age of 2-2.5~Myr and an initial mass of 
$\sim$110M$_{\odot}$.  The  X-ray properties of  \object{CXOGC J174536.1-285638} also 
 resemble those of 3 of the 4 X-ray detected WN9h stars within the Arches   and in turn 
other very massive WNLh colliding wind binaries,  of which 
  WR25 forms an almost identical `twin'. Simple analytical arguments demonstrate
consistency between the X-ray emission and  a putative  WN9h+mid O~V-III binary, 
causing us to favour such a scenario over an accreting binary. However, we may not 
exclude a high mass X-ray binary interpretation, which, if correct,  would provide a 
unique insight into the (post-SN) evolution of extremely  massive  stars. 
Irrespective of the nature of the secondary, \object{CXOGC J174536.1-285638} adds to the  
growing list of known and candidate WNLh binaries. Of the subset of WNLh stars
 subject to a radial velocity survey, we find a lower limit to 
the  binary fraction of $\sim$45\%; of interest for studies of 
massive stellar formation, given that they currently
 possess the highest dynamically determined masses of any type of star.
}{}


\maketitle

\section{Introduction}

Near-IR observations of the Galactic centre have demonstrated that it hosts a 
large  population of high mass stars, predominantly located  within three
 young (2-6~Myr), massive ($>10^4$M$_{\odot}$) clusters in the 
central $\sim$50pc - the Arches, Quintuplet and Central cluster (e.g. Krabbe et al. \cite{krabbe}, Nagata et al. \cite{nagata}, Cotera et al. \cite{cotera}, Figer et al. \cite{figer95}). 
Recently, the combination of radio, X-ray and near-mid IR data
has revealed an additional  population of apparently isolated (candidate) massive stars throughout this region
 (e.g. Mauerhan et al. \cite{mauerhan}). A full understanding of the processes governing
 star formation in the extreme environment of  the Galactic   centre therefore requires an explanation 
for their properties and origin - are they remnants of clusters disrupted by tidal forces or interaction 
with molecular clouds, the result of dynamical or SNe kick ejection from natal clusters or did they form {\em in situ} 
in a non clustered environment?

Identification of such stars has been facilitated by the production of deep X-ray catalogues of the Galactic Centre (GC; 
Muno et al. \cite{muno06a}, \cite{muno09}), revealing over 9000 discrete point sources in the central 
2$^{\rm o} \times 0.8^{\rm o}$. Of these the majority are expected to be low mass systems, most likely  Cataclysmic Variables, 
but a  subset of the brighter, variable sources are expected to be either Colliding  Wind Binaries (CWBs) or 
High Mass X-ray Binaries (HMXBs). Cross correlation of these catalogues with near-IR imaging and subsequent spectroscopy 
 has revealed a number of these sources to be  identified with candidate high stellar mass counterparts 
(Muno et al. \cite{muno06b}, Mauerhan et al. \cite{mauerhan})
Critically, the star formation rate inferred for the GC suggest a statistically significant number of HMXBs (Muno et 
al. \cite{muno06a} and refs. therein) should be present and detectable, allowing the physical 
 assumptions of population synthesis modeling (such as the SNe kick velocity) to be tested. Therefore, accurate 
classification of these systems is invaluable in order to confirm their nature and the origin of the X-ray emission. 

Mikles et al. (\cite{m06}, \cite{m08}; henceforth M06, M08) report the discovery of an emission line star associated
 with the bright X-ray source \object{CXOGC J174536.1-285638} (abbreviated to CXO J1745-28). The X-ray spectrum is
dominated by strong Fe\,{\sc XXV} emission, with the 
luminosity 
(L$_x \sim 1.1\times10^{35}$ergs$^{-1}$) and hard nature of the  emission  ($kT \sim 
0.7^{+0.1}_{-0.1}+4.6^{+0.7}_{-0.7}$keV; M06) both arguing for a binary interpretation. However,  the nature of the 
system (CWB or HMXB) remains obscure, even after the discovery of a 189$\pm$6~d periodicity in the X-ray flux
(M08). This is largely due to the uncertainty regarding the  properties of the primary and hence in
 this study we  present a detailed analysis of the near-IR spectroscopy presented by M06 in order to provide
 accurate stellar parameters for the system primary and hence address the nature of the system. In Sect. 2 we
 present the results of this non-LTE analysis, discuss the nature of the system in Sect. 3 and summarise
our results in Sect. 4.  

\begin{table*}
\caption[]{UKIDSS near-IR photometry, extinction and absolute magnitude determinations for
an assumed distance of 8 kpc to  CXO J1745--28 (Reid \cite{1993}).
}
\label{photom}
\begin{center}
\begin{tabular}{llllllllllll}
\hline
Star  & K & J--K & H--K & 
(J--K)$_0$ & (H--K)$_0$ & 
$A_{\rm K}^{J-K}$ & $A_{\rm K}^{H-K}$ & 
$A_{\rm K}$ & DM & $M_{\rm K}$ \\
\hline
CXO J1745--28 & 10.40 & 5.07 & 1.78 & --0.04 & 0.01 & 3.42 & 3.22 & 3.32 & 
14.52 
& --7.51 \\
\hline
\end{tabular} 
\end{center}
\end{table*}

\begin{figure}
\begin{center}
\includegraphics[width=\columnwidth]{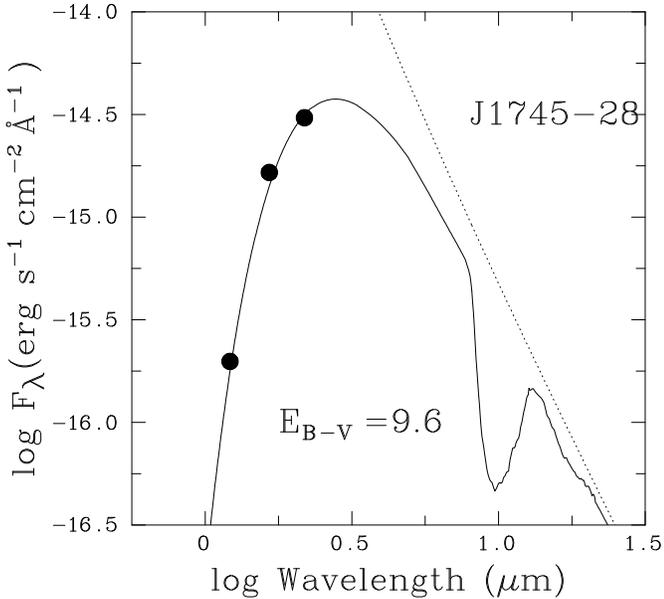}
\caption[]{Reddened model spectral energy distribution (SED) of CXO J1745--28
(solid line, $E_{B-V}$=9.6) and intrinsic SED (dotted line) overlaid upon 
IR photometry from UKIDSS}\label{SED}
\end{center}
\end{figure}

\section{Observations and analysis}

We have carried out a non-LTE 
spectroscopic analysis of the near-IR 
spectroscopy of CXO J1745--28 obtained by M06 using IRTF/SpeX
(Rayner et al. \cite{r03}) and CMFGEN (Hillier \& Miller \cite{hil98}).

\subsection{IRTF/SpeX spectroscopy}

Short-wavelength, cross-dispersed spectroscopy of CXO J1745--28
was obtained by M06 on 1 Jul 2005 using IRTF/SpeX, covering the 
JHK bands  at a spectral resolution of $R\sim$1200. M06 describe
the data reduction process and spectral features which we shall
not repeat here, except that the broad emission feature at 1.77$\mu$m
attributed to He\,{\sc ii} (19-8) 1.772$\mu$m is in error.
He\,{\sc ii} is expected to make a minor contribution to this
feature, on the basis of negligible emission from He\,{\sc ii} (23-8)
1.658$\mu$m (21-8 is blended with He\,{\sc i} 1.700$\mu$m). 
The primary component is not known. 

\begin{figure}
\resizebox{\hsize}{!}{\includegraphics[angle=270]{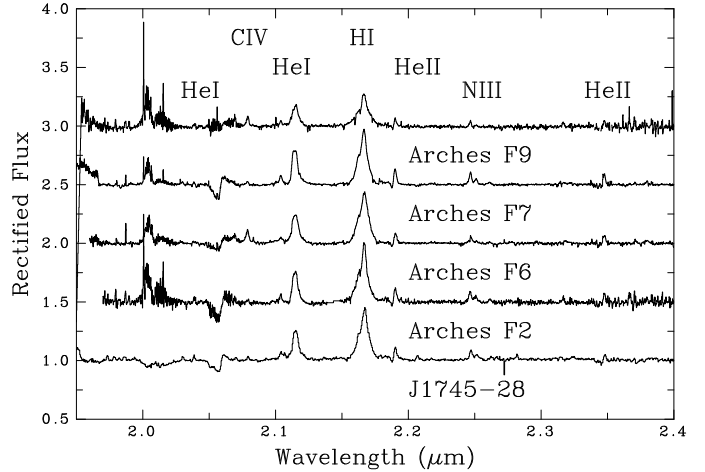}}
\caption{K-band spectroscopy of CXO J1745-28 from M06 together with
morphologically similar WN9h stars from the Arches cluster 
(F2, F6 and F7) plus the weaker lined WN9h star F9 which is X-ray bright,
as are F6 and F7. The VLT/SINFONI Arches spectroscopy is from
Martins et al. \cite{martins08}). }\label{arches}
\end{figure}

Crowther et al. (\cite{PAC}) discuss near-IR spectral classification of 
Wolf-Rayet stars from which we assign a WN9 subtype on the basis of 
He\,{\sc ii}/Br$\gamma$ = 0.06$\pm$0.01. We may refine this subtype in 
view of its close morphological match to near-IR spectroscopy of 
HDE\,313846 (WR108) from Bohannan \& Crowther (\cite{bc99}). HDE\,313846 
is a WN9ha star, for which the `h' indicates the presence of significant 
hydrogen from the Pickering-Balmer decrement at visual wavelengths and the 
`a' represents intrinsic absorption lines in high Balmer lines. Analogous 
`h' diagnostics are available in the near-IR, such as the hydrogen 
Brackett series and corresponding He\,{\sc ii} (n--8) transitions in 
CXO J1745--28. We therefore refine the subtype of CXO J1745--28 to WN9h. The 
morphological similarity between CXO J1745--28 and HDE\,313846 suggests both 
are WN9ha subtypes, although no unambiguous diagnostic is available from 
low resolution near-IR spectroscopy. Other examples of WN9ha stars are 
known, such as HD\,152408 (Bohannan \& Crowther \cite{bc99}) which had 
earlier been classified as an Ofpe star by Walborn (\cite{w82}), 
reflecting their proximity to the Wolf-Rayet and Of boundaries. Recently, 
many other examples of WN7--9h stars have been identified in the Arches 
cluster from near-IR spectroscopy (Martins et al. \cite{martins08}). Of 
these, F2 and F7 possess essentially identical K-band spectral 
morphologies to CXO J1745--28, while F6 only differs in its stronger C\,{\sc 
iv} 2.070, 2.079$\mu$m features, as shown in Fig.~\ref{arches}. We also 
include one other WN9h star F9 for comparison since this is a strong X-ray 
source (as are F6 and F7), although its emission line spectrum is 
somewhat 
weaker.

\begin{figure}
\resizebox{\hsize}{!}{\includegraphics[angle=0]{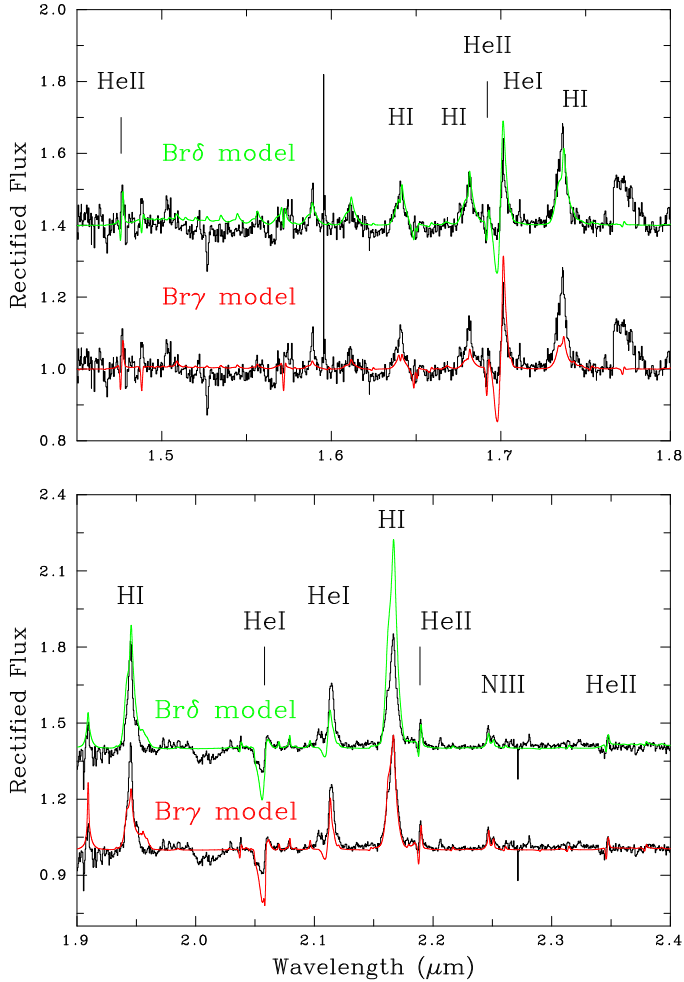}}
\caption{H and K band IRTF/SpeX observations of 
\object{CXO J1745-28} from M06 (black)
with our Br$\gamma$-derived (red, $X_{\rm H}$=25\%) and Br$\delta$-derived
(green, $X_{\rm H}$=50\%) synthetic spectra overplotted.}\label{fit}
\end{figure}

We have used photometry of CXO J1745--28 from the fourth data release (DR4) of 
UKIDSS together with intrinsic JHK colours predicted by our spectroscopic 
analysis (presented in Sect.~\ref{analysis}), from which $K$-band 
extinctions, $A_{\rm K}$, may be obtained using the extinction 
relations from Indebetouw et al. (\cite{i05}). The UKIDSS project is 
defined in Lawrence et al (\cite{ukidss}). UKIDSS uses the UKIRT Wide 
Field Camera (WFCAM; Casali et al, \cite{casali07}). The photometric 
system is described in Hewett et al (\cite{hewett08}), and the calibration 
is described in Hodgkin et al. (\cite{hodgkin09}).

We favour UKIDSS photometry to 2MASS, owing to a late-type source which 
is blended with CXO J1745--28 in 2MASS datasets (see M08). Our 
derived extinction of $A_{\rm K}$ = 3.32 mag for CXO J1745--28 corresponds to 
$E_{B-V}$=9.6 or $A_{\rm V}$=30 mag, assuming a standard Galactic 
extinction law. For an assumed distance of 8 kpc to the Galactic Centre 
(Reid \cite{1993}), we obtain an absolute $K$-band magnitude of $-$7.51 
for 
CXO J1745--28. In Fig.~\ref{SED} we present the spectral energy distribution 
(SED) of our J1745--28 model, reddened according to the IR extinction law 
presented in Morris et al. (\cite{morris}), together with UKIDSS photometry. 
Spitzer IRAC photometry is not available, although the theoretical model 
suggests [3.6]=8.28, [4.5]=7.66, [5.8]=7.24 and [8] = 7.11 for J1745--28. 
Our derived extinction agrees well with M06 who estimated A$_{\rm V}$=29 
mag by simply adopting $(H-K)_{0}$=0.0 mag.

For comparison, WN9h stars within the Arches cluster possess HST/NICMOS 
F205W magnitudes in the range 10.45 (F1) to 11.2 (F2) based upon 
photometry from Figer et al. (\cite{figer2002}). These correspond to 
absolute F205W magnitudes of --6.4 to --7.2 mag for a uniform 
K-band extinction of $A_{\rm K_{s}}$ = 3.1 mag (Kim et al. \cite{kim06}) 
and  (identical) distance of 8 kpc. As such, one would expect that 
CXO J1745--28 to be amongst the most luminous of the WN9h stars in the 
Galactic Centre region.

\begin{table*}
\begin{center}
\caption{Derived physical and wind parameters for \object{CXO J1745-28} (WN9h) 
for CWB and HMXB scenarios}\label{params}
\begin{tabular}{llllllllllll}
\hline
Scenario   & Model & $T_{\ast}$ & $R_{\ast}$ & $T_{\rm eff}$ & $\log L_{\ast}$ & $v_{\infty}$ & 
$\log$ dM/dt & $M_{\rm K_{s}}$ & $X_{H}$ & $X_{\rm He}$ & $X_{\rm N}$ \\
           & & kK & $R_{\odot}$ & kK & $L_{\odot}$ & km\,s$^{-1}$ & $M_{\odot}$\,yr$^{-1}$ &
mag & \% & \% & \% \\
\hline
CWB  & Br$\gamma$ & 32.0 & 43.1 & 30.9 & 6.24 & 1350 & --4.46 & --7.42 & 25 & 74 & 1.2 \\
     & Br$\delta$ & 32.5 & 41.8 & 30.5 & 6.24 & 1350 & --4.40 & --7.42 & 50 & 49 & 1.2 \\
HMXB & Br$\gamma$ & 32.0 & 44.9 & 30.9 & 6.28 & 1350 & --4.43 & --7.51 & 25 & 74 & 1.2 \\
     & Br$\delta$ & 32.5 & 43.5 & 30.5 & 6.28 & 1350  &--4.37  & --7.51 & 50   & 49   & 1.2    \\ 
\hline
\end{tabular}
\end{center}
\end{table*}

\subsection{Atmospheric code}

The non-LTE atmosphere code CMFGEN solves the radiative transfer equation 
in the co-moving frame, 
under the additional constraints of statistical and radiative equilibrium.

Since CMFGEN 
does not solve the momentum equation, a density or velocity structure is 
required. For the supersonic part, the velocity is parameterized with a 
classic $\beta$-type law. This is connected to a hydrostatic density 
structure at depth, such that the velocity and velocity gradient match at 
the interface. The subsonic velocity structure is defined by a 
corresponding  $\log g = 3.25$ fully line-blanketed plane-parallel TLUSTY 
model (Lanz \&  Hubeny \cite{lh03}). 

CMFGEN incorporates line blanketing through a super-level 
approximation, in which atomic levels of similar energies are
grouped into a single super-level which is used to compute the atmospheric
structure. Our atomic model is similar to that adopted by 
Crowther et al. (\cite{c02}), including ions from H\,{\sc i}, He\,{\sc 
i-ii}, C\,{\sc iii-iv}, N\,{\sc iii-v}, O\,{\sc iii-vi}, Si\,{\sc iv}, 
P\,{\sc iv-v}, 
S\,{\sc iv-vi} and Fe\,{\sc iv-vii}. By number, the main 
contributors to line blanketing are Fe\,{\sc iv-v}. In addition, extended 
model atoms of C\,{\sc iii}, N\,{\sc iii} and O\,{\sc  iii} were included 
since each contribute to the 2.112$\mu$m emission feature.

We have assumed a depth-independent Doppler profile for all lines when 
solving for the atmospheric structure in the co-moving frame, while in the 
final calculation of the emergent spectrum in the observer's frame, we 
have adopted a uniform turbulence of 50 km\,s$^{-1}$. Incoherent electron 
scattering and Stark broadening for hydrogen and helium lines are adopted. 

With regard to wind clumping, this is incorporated using a radial
dependent volume filling factor,  $f$, as described in Hillier et al. 
(\cite{h03}), with a typical value of  $f$=0.1 resulting in a reduction in 
mass-loss rate by a factor of $\sqrt{(1/f)} \sim 3$.

\subsection{Analysis}\label{analysis}

We derive the stellar temperature CXO J1745--28 using diagnostic He\,{\sc i} 
2.058$\mu$m, 1.700$\mu$m, He\,{\sc ii} 2.189$\mu$m, 1.692$\mu$m together 
with Br$\gamma$ for the mass-loss rate, hydrogen content and velocity 
structure. The P Cygni He\,{\sc ii} 2.189$\mu$m profile naturally arises 
in the stellar wind of the star and {\em does not} require the presence of 
an additional source of ionisation within the system, such as a compact 
companion (e.g. M06). Stellar temperatures, $T_{\ast}$, correspond to a 
Rosseland  optical depth of 20, which are typically (up to) a few thousand 
degrees  higher than effective temperatures, $T_{2/3}$, relating to 
optical depths of 2/3 in such stars.

We have estimated a terminal wind velocity of 1350 km\,s$^{-1}$ from 
He\,{\sc i} 2.058$\mu$m from which a (slow) velocity law of exponent 
$\beta$=1.5 is used for the supersonic velocity structure. $\beta$ = 0.8
was adopted by Martins et al. (\cite{martins08}), for which similar
synthetic spectra are predicted, except for somewhat weaker emission 
and absorption lines. A measure of the nitrogen mass fraction is obtained 
from the weak N\,{\sc iii} 2.247,  2.251$\mu$m doublet since the stronger 
feature at 2.112$\mu$m is blended  with He\,{\sc i}, C\,{\sc iii} and 
O\,{\sc iii}, while C\,{\sc iv}  2.070, 2.079$\mu$m allows an estimate of 
the carbon content. We adopt  solar  values for all other metals
(e.g. Cox \cite{cox}, with the exception of oxygen for which the value given 
by  Asplund et al. \cite{asplund} was used). 
Spectroscopic fits to  IRTF/SpeX observations are presented in 
Fig.~\ref{fit}. 

Overall the agreement between predicted line profiles and observations is 
satisfactory, although He\,{\sc i} 2.058$\mu$m and 1.700$\mu$m P Cygni 
absorption are too strong in the synthetic spectrum ($\beta$ = 0.8 does provide an 
improved match to 2.058$\mu$m)  and it is apparent  that the high members 
of the hydrogen Brackett series  are predicted to be too weak in emission 
for the adopted $X_{\rm H}$=25\%  by mass (He/H=0.7 by number). This model 
is referred to as the Br$\gamma$  model in Table~\ref{params}, although it 
should be emphasised that  He\,{\sc i} lines within the 7--4 set of 
transitions contribute  $\sim$30\% of the equivalent width for this 
feature. 

In  order to better reproduce the strength of the higher Brackett 
series,
a significantly higher hydrogen content of $X_{\rm H}$=50\% by mass 
(He/H=0.25 by number) is required, which is also presented in Fig.~\ref{fit}.
Model parameters for this, our Br$\delta$ model, are also presented in 
Table~\ref{params}.  Consequently, one should caution against the use of 
a solitary diagnostic in view of the significant difference resulting 
from other  diagnostics even for cases as simple as a single hydrogen 
series. 

In both cases, we  estimate a nitrogen mass fraction of 1.2\% for 
CXO J1745--28 from the N\,{\sc iii} 2.247, 2.251$\mu$m features which should 
be 
reliable to $\pm$50\%. Carbon is poorly  constrained, although $X_{\rm 
C}$=0.03\% by mass reproduces the weak C\,{\sc iv} 2.070, 2.079$\mu$m 
features satisfactorily.

In view of the uncertain nature of the companion to the WN9h star (Sect. 3), we have 
estimated its stellar properties under two assumptions; either the near-IR 
spectrum of the Wolf-Rayet star is (weakly) diluted by the continuum of a 
massive companion in the case of a CWB system or it suffers negligible
contribution from the compact companion in the case of a HMXB. Here, we 
adopt a mass ratio $q$ = 0.4, which is consistent with $q\geq$0.2 
favoured by M08 for a sole dominant IR source. For a current WN9h mass of 
80 $M_{\odot}$ (see below), we shall use 30  $M_{\odot}$ for the companion 
mass, namely an O6 dwarf, O7 giant or O9 supergiant from the Martins et 
al. (\cite{martins05}) O star calibration. On the basis of the companion wind 
properties, discussed in Sect.~\ref{binary}, we shall adopt O7\,III for 
the putative companion, for which $M_{\rm K}$ = --4.76 (Martins \& Plez 
\cite{mp06}).  For a systemic absolute magnitude of $M_{\rm K}$ = --7.51
mag we obtain $M_{\rm K}$ = --7.42, i.e. the light ratio of the O to WR 
star is 0.09 in the K-band (0.10 in the J and H bands). Derived stellar 
parameters for the WN9h star in these scenarios are presented in 
Table~\ref{params}.

\subsection{Comparison with Arches cluster members}

\begin{figure}
\resizebox{\hsize}{!}{\includegraphics[angle=0]{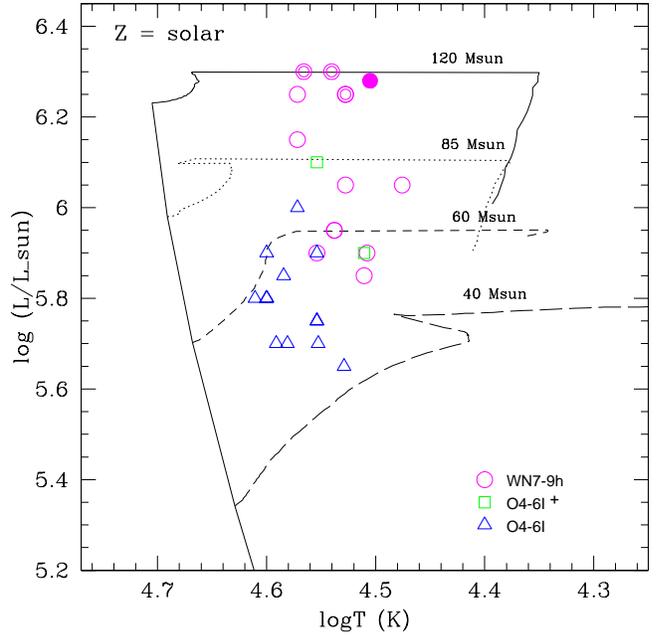}}
\caption{HR diagram indicating the position of \object{CXOGC J1745-28} 
(filled symbol) and the WN7-9ha 
and O4-6I members of the Arches cluster (open symbols; Martins et al. 
\cite{martins08}), with the luminosity of Arches members revised downwards by 0.05dex 
as described in Sect. 2.4. The  Geneva evolutionary tracks for rotating (V$_{rot}=300$kms$^{-1}$)
40-120M$_{\odot}$ stars at solar metallicity are overplotted (Meynet \& Maeder 
\cite{meynet}). The Arches stars F6, 7 \& 9 (CWBs based on their X-ray properties, 
see Sect 3)  are indicated by the double rings.}\label{hrd}
\end{figure}

The position of CXO J1745--28 on the H-R diagram is presented in 
Fig.~\ref{hrd} together with O supergiant and WN7--9h members of the 
Arches cluster from Martins et al. (\cite{martins08}) and solar 
metallicity, rotating (300 km\,s$^{-1}$) theoretical models from Meynet \& 
Maeder (\cite{meynet}). Note that the stellar luminosities of the Arches 
members have been downward corrected by 0.1 dex due to an error in the 
original study (Martins, priv. comm.), but re-adjusted upward by 0.05 dex
for consistency with our adopted distance.\footnote{Martins et al. 
(\cite{martins08}) adopted a distance of 7.6 kpc to the Arches cluster, 
a K-band extinction of 2.8 mag and F205W photometry from Figer et al. 
(\cite{figer2002}).}
 Arches WN9h members F2, F6 and F7 
are of particular interest in view of their spectroscopic 
similarities to CXO J1745--28 (recall Fig.~\ref{arches}), for which Martins et 
al. (\cite{martins08}) obtained stellar temperatures of 34kK (versus 
32kK here for CXO J1745--28). However, $K$-band bolometric corrections, 
BC$_K$, for the Arches members remain significantly larger (BC$_{\rm 
F205W} \sim -$4 
mag) than our estimate for J1745--28 (BC$_K \sim -$3.5 mag), even after 
the 0.1 dex correction in stellar luminosity. If we turn to abundance
estimates, Br$\gamma$-derived  surface hydrogen contents in the range 40\% 
$\leq X_{H} \leq$ 55\% are obtained for F2, F6 and F7, versus 25\% for 
CXO J1745--28, while the nitrogen mass fraction  obtained for CXO J1745--28 
is also at the lower end of the range observed for the Arches members; 1.2\% versus
1.1--2.8\% respectively.

We infer an initial (current) mass of $\sim110$ (80) $M_{\odot}$ and a 
current age of 2--2.5 Myr for CXO J1745--28, such that it is likely a very 
massive star that is slowly evolving away from the hydrogen burning main 
sequence. As such, J1745--28 is not a classical helium-burning Wolf-Rayet 
star.

\begin{figure}
\resizebox{\hsize}{!}{\includegraphics[angle=0]{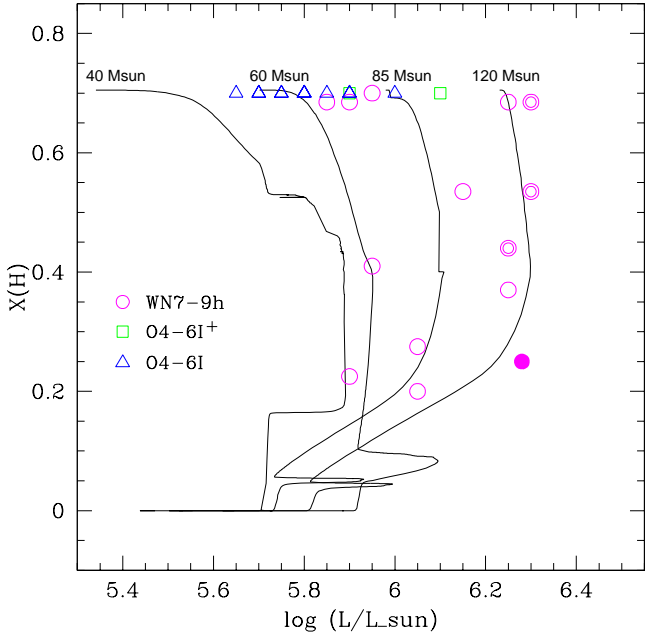}}
\caption{Comparison of the Hydrogen mass fraction as a function of luminosity
 for \object{CXOGC 1745-28} and the stellar population of the  Arches cluster, with 
symbols having the same meaning as in Fig.~\ref{hrd}. To enable a direct  comparison to 
the Arches population the results from the Br$\gamma$ model are presented  - $X_{\rm H}=50$\%
by mass for the Br$\delta$ model (Sect. 2.3).  Again, Geneva evolutionary 
tracks for rotating (v$_{rot}=300$km~s$^{-1}$) 40-120M$_{\odot}$ stars at solar metallicity have been 
overplotted (Meynet \& Maeder \cite{meynet}).}\label{h}
\end{figure}

Predicted surface hydrogen mass fractions at this early stage are $X_{\rm 
H}$=40\% for models initially rotating at 300 km\,s$^{-1}$, with lower 
depletions of $X_{\rm H}$=65--70\% for equivalent non-rotating models. In 
Fig.~\ref{h} we compare the Br$\gamma$-derived surface hydrogen content 
for CXO J1745--28  with Arches cluster members from Martins et al. 
(\cite{martins08}) and the  predicted surface hydrogen from Meynet \& 
Maeder (\cite{meynet}) models.  Recall that the hydrogen content for 
CXO J1745--28 depends upon the diagnostic  hydrogen line used, either $X_{\rm 
H}$=25\% from Br$\gamma$ (the same diagnostic as used by Martins et al. 
\cite{martins08}) or $X_{\rm H}$=50\% for higher Brackett lines.

\section{The nature of CXO J1745-28 - CWB or HMXB?}\label{binary} 

Our classification of the primary in \object{CXO J1745-28} as a WN9h star  permits a more accurate comparison
to the properties of both single stars and known HMXBs and CWBs. 
The ratio of X-ray to bolometric luminosity of \object{CXO J1745-28} - log(L$_x$/L$_{bol}$)$\sim$-4.8 - is 
significantly in excess of that observed for both single O and WR stars - log(L$_x$/L$_{bol}$)$\sim$-7 - where the 
emission is thought to arise in shocks embedded in the stellar wind.
Recent work has shown that a number of O stars have measureable magnetic fields
(Donati et al. \cite{donati02}, \cite{donati06}; Hubrig et al. \cite{hubrig}, Bouret et al. \cite{bouret}), thus 
raising the possibility that the magnetically confined wind shock model  of Babel \& 
Montmerle et al. (\cite{babela}, \cite{babelb}) might be applicable and lead to the production of significant  X-ray emission. 
In Table 3 we summarise the X-ray  properties of these stars as well as the Of?p star \object{HD 108}  (motivated by the 
fact that the other three examples of Of?p stars all have detectable magnetic fields).

While these stars appear  to show excess X-ray emission with respect to field O and WR stars (Table 3; N\'aze et al. 
\cite{naze07}), the ratio of X-ray to bolometric luminosity is over an order of magnitude lower than 
\object{CXO J1745-28}, with the  WN9ha star HD 152408 (= WR 79a) being significantly fainter still.
 Moreover, while the X-ray spectra of the magnetically active stars are also 
consistent with a multitemperature model, both cool and hot components are 
systematically cooler than found for \object{CXO J1745-28} (Rauw et al., \cite{rauw02}, G\'agne 
et al. \cite{gagne}, Naze et al. \cite{naze04}, \cite{naze07}, \cite{naze08b}). Finally, these stars demonstrate
spectral variability attributed to an asymmetric circumstellar environment caused by  the 
entrainment  of the stellar wind by the  magnetic field (e.g. N\'aze et al. \cite{naze08}). However, spectroscopy of 
\object{CXO J1745-28}  indicates there is no evidence for a departure from spherical symmetry, while no variability has 
currently been observed (Mikles et al. \cite{m06}, \cite{m08}). 
Therefore, we conclude that the X-ray emission in \object{CXO J1745-28} most likely arises 
as a result of binarity.

Assuming  the 189~d periodicity is orbital and a canonical  
accretion efficiency $\epsilon \sim0.1$, the wind properties imply a compact companion mass of 
 $\sim$5M$_{\odot}$  for a HMXB interpretation (Eqn 10; M08).  Three HMXBs with  moderately
 evolved very massive companions are known;
4U1700-37, OAO1657-415 \& GX301-2\footnote{4U1700-37 - O6.5 Iaf$^+$, P$_{orb}$=3.412d (Clark et al. \cite{c02});  OAO1657-415 - 
WN9-11h, P$_{orb}$=10.4d (Mason et al. \cite{mason}); GX301-2 - B hypergiant, P$_{orb}$=41.5d (Kaper et al. \cite{kaper}).}.
All are more X-ray luminous than \object{CX J1745-28}; as expected, given 
their smaller orbital separations. However their X-ray spectra - a power law with high energy cutoff, with no 6.7keV 
Fe\,{\sc XXV} line (White et al. 
\cite{white}, La Barbera et al. \cite{lab}, Audley et al. \cite{audley})- are significantly different\footnote{The accretors in
GX301-2 and OAO1657-415 are pulsars, while the nature of 4U1700-37 is uncertain; however for the persistent  Black Hole 
systems such as Cyg X-1 a composite  multitemperature disc+power law spectrum is typically observed 
(e.g. Cui et al. \cite{cui}), which also appears to differ from \object{CXO J1745-28}.}.

Motivated by the similarity in X-ray properties of \object{CXO J1745-28} to the CWBs \object{$\eta$ Carinae} 
(e.g. Pittard \& Corcoran \cite{pittardc}) and Cyg OB2\#8A (DeBecker et al. \cite{deB}), we summarise 
the  properties of  WNLh CWBs in the Galaxy and Large Magellanic Cloud in Table 4. 
 While the X-ray luminosity of the sample is  observed to range  over two orders of magnitude, \object{WR25}, BAT99-112  
\& 116  are directly comparable to \object{CXO J1745-28}.  Likewise, where it may be  determined, the emission from
 WNLh CWBs is significantly harder than $kT\sim$0.6~keV expected for shock emission in a single stellar wind (e.g. Oskinova 
\cite{oskinova}), and for both WR20a and WR25 a two component fit is required, as found for \object{CXO J1745-28}. 
In addition to the comparable
 X-ray luminosity and  spectral components, WR25 is of particular interest given  the presence of strong  Fe\,{\sc XXV} emission, 
variable X-ray flux (Pollock \& Corcoran \cite{pollock06}) and 
a spectroscopically determined orbital period of 207.8$\pm$0.3d (Raassen et al. \cite{raassen}).  

We consider these striking  similarities as strong evidence for a CWB interpretation for \object{CXO J1745-28}. 
However, one would like to identify further examples of such systems in order to cement such an interpretation. 
Intriguingly, 3 of the 4 WN8-9h stars within the Arches (Table 4) which have X-ray detections 
also demonstrate similarly strong   Fe\,{\sc XXV} emission to \object{CXO J1745-28} (the fourth, F2, being too faint to
determine if this is the case). Spectral fits to these stars 
 require a hard component for either a one (Wang et al. \cite{wang}) or two (Yusef-Zadeh et al.   \cite{yz02}) temperature model,
with the resultant  luminosities - 10$^{33}$ergs$^{-1}$ and  10$^{35}$ergs$^{-1}$ respectively - also directly comparable to 
that of \object{CXO J1745-28} under the same model assumptions.  Given the similarity of the stellar+wind properties of 
the Arches members to \object{CXO J1745-28} (Sect 2.4 and Figs. 2, 4 \& 5) we therefore conclude that these are 
physically $\sim$identical systems.  This  in turn provides additional support for a CWB interpretation. The 
WN7-9ha stars within the Arches are the most massive stars present, and are at an age when it is not expected that
 {\em any} stars will have been lost  to SNe (Table 5). Consequently, it is not apparent that an HMXB interpretation is 
possible on evolutionary grounds, still less since one would also have to conclude that the  post SNe orbital configurations 
for \object{CXO J1745-28} and  Arches F6, 7 \& 9 would have to be $\sim$identical in order to reproduce their common  X-ray 
properties.

Under the hypothesis that \object{CXO J1745-28} is a CWB can we draw any conclusions as to the nature of the hitherto 
unseen companion?  Given the absence of a radial velocity (RV) curve tailored hydrodynamical simulations are clearly
premature but it is possible to estimate bulk properties for the secondary. Initially, the strong shock jump 
conditions ($kT=(3/16)$\={\em m}$v^2$,  where \={m} is the mean particle mass - 10$^{-27}$kg for solar abundances (Stevens et al. \cite{stevens})  - and the other symbols have their usual meanings) 
gives $v\sim2100$kms$^{-1}$. This is  consistent
with the mean  terminal velocity of mid-late O stars (Prinja et al. \cite{prinja}), but not with
the W7-9h stars within the Arches (Martins et al. \cite{martins08}), implying that a system consisting of twin 
WNLh stars such as WR20a is not viable.

We may use the analytical expressions from  Pittard \& Stevens (\cite{pittards}) to estimate  the X-ray flux for a
 representative WN9h+mid O star binary  to test the consistency of such an hypothesis. Assuming M$_{\rm WR}\sim80$M$_{\odot}$
 (Sect. 2.4 and Fig. 4), M$_{\rm O}  \sim30$M$_{\odot}$ and an 189d period
  we derive a binary separation of $\sim5\times10^{13}$cm. We first define the momentum ratio 
of the two winds $\eta$=\.{M}$_{\rm O} v_{\rm O}$/\.M$_{\rm WR} v_{\rm WR}$. Then the  kinetic  power
processed for {\em each} star is given by $L=0.5$\.{\em M}$v^2 \Xi$, where the fractional wind kinetic power $\Xi$ 
is a function of $\eta$. With the properties of the  primary from Table 2, $v_{\rm O} \sim 2100$kms$^{-1}$ and \.{M} in the range
2-20$\times 10^{-7}$ M$_{\odot}$yr$^{-1}$ we find $\eta$=0.01-0.1 and corresponding values for $\Xi_{\rm WR}$=0.004-0.033
and $\Xi_{\rm O}$=0.564-0.403 (Pittard \& Stevens \cite{pittards}). Then we may estimate the intrinsic X-ray luminosity for 
each star as $L_x=0.5$\.{\em M}$v^2 \Xi /\chi$, where the cooling efficiency, $\chi$, takes the conversion 
efficiency of kinetic wind power into radiation into account, and is given by $\chi = {v_8}^4 d_{12}/$\.{\em M}$_{-7}$ (where $v_8$ is wind velocity in units of 1000kms$^{-1}$, 
$d_{12}$ is the distance from the star of the contact discontinuity in units of 
10$^{12}$cm and \.{M}$_{-7}$ the mass loss rate in units of 10$^{-7}$M$_{\odot}$yr$^{-1}$). Note that if $\chi<$1.0 it is
 set to 1.0 for  the calculation of X-ray luminosity since one may not radiate more energy than one has as input (Pittard, piv. comm. 2009). 

We estimate $d_{12}$ from the results presented in Pittard \& Stevens (\cite{pittards}) and then substituting the appropriate 
values  into the above relationships we find L$_{\rm X,WR} \sim 8\times10^{34}$ergs$^{-1}$ and L$_{\rm X,O} 
\sim9\times10^{34}$ and 3$\times10^{33}$ergs$^{-1}$ for \.{M}=20-2$\times10^{-7}$M$_{\odot}$yr$^{-1}$. Therefore 
 we consider the X-ray luminosity of \object{CXO J1745-28} to be broadly consistent with that expected from 
 a binary composed of a massive WN9h primary and a less evolved $\sim$mid  O star secondary in a 189d period orbit, 
although a full hydrodynamical simulation would be required to extract  accurate predictions for the properties of the 
secondary (c.f. Pittard \& Corcoran \cite{pittardc}).

\begin{table*}
\begin{center}
\caption[]{Summary of the properties of CXO J17545-28 in comparison to magnetic O and WR and Of?p stars}
\begin{tabular}{lccccc}
\hline
\hline
ID & Spec. & log(L$_{bol}$) & log(L$_x$) & log(L$_x$/L$_{bol}$) & Ref. \\
   & Type  &  (erg s$^{-1}$)& (erg s$^{-1}$) & & \\
\hline
CXO J1745-28 & WN9h      & 39.8 & 35.0 & -4.8 & 1, this study \\
             &           &      &      &      &                    \\
$\theta^1$ Ori C  & O7 V & 39.0   & 33.0 &   -6.0 & 2,3,4 \\
$\zeta$ Ori A     & O9.7Ib & 38.8  & 32.1     & -6.7  & 5,6  \\      
HD 108  & O4f?p$\leftrightarrow$O8.5fp & 39.2 & 33.1 & -6.1 & 2,3 \\
HD 148937 & O5.5-6f?p & 39.3 & 33.3 & -6.0 & 3,7 \\
HD 152408 (=WR 79a)           & WN9ha& 39.6   & 31.9 & -7.7 & 8    \\ 
HD 164794    &  O4 V((f))& 39.5   & 33.1  & -6.4 &  9 \\ 
HD 191612    & O6.5f?pe$\leftrightarrow$O8fp & 39.1 & 32.9 & -6.1 & 2 \\
\hline
\end{tabular}
\end{center}
{Note that no magnetic field determinations have been made for HD108, while HD 164974 (=9 Sgr) is a suspected 
CWB. References for this table are:
(1) Mikles et al. (\cite{m06});
(2) N\'aze et al. (\cite{naze07});
(3) N\'aze et al. (\cite{naze08});
(4) Gagn\'e et al. (\cite{gagne});
(5) Raassen et al. (\cite{raassenmag});
(6) Bouret et al. (\cite{bouret});
(7) N\'aze et al. (\cite{naze08b})
(8) Oskinova (\cite{oskinova});
and 
(9) Rauw et al. (\cite{rauw02}).}
\end{table*}

\begin{table*}
\begin{center}
\caption[]{Summary of the stellar, orbital and, where determined, the 
 X-ray properties of confirmed or candidate  massive WNLh binary systems in the Galaxy and LMC}
\begin{tabular}{lcccccccc}
\hline
\hline
ID         & Spec.  & log(L$_{WR}$)  & M$_{WR,dynamical}$&P$_{\rm orb}$   &  L$_x$               & $kT$  & FeXXV & References\\
           & Type   & (L$_{\odot}$)  & (M$_{\odot}$)     & (days)         &(10$^{33}$ergs$^{-1}$)& (keV) &  (Y/N)  &         \\
\hline
CXO J1745-28&WN9h +&   6.24        &      -             & 189$\pm$6     & 110                  & 0.7+4.7 &   Y   & 1,2  \\ 
           & mid-O? &    -           &      -             &               &                      &         &       &      \\
           &        &    -           &      -             &               &                      &         &       &      \\
BAT99-116 (Mk34)   & WN5h    &    -          &      -             & RV var.       & 240                  &   3.9   &  -    & 3,4 \\
WR25 (HD 93162)      & WN6h + ?&               &      -             & 207.8$\pm$0.3 &  130$\pm$10          & 0.7+2.8 & Y     &  5,6  \\  
BAT99-112 (R136c) &  WN5h   &    -          &      -             & RV var.       & 110                  &   3.0   &  -    & 3,7 \\ 
\hline     
NGC3603 C  &WN6ha + ? & 6.1          &      -              & 8.89$\pm$0.01 &  $>$40               & - &  -   &  8,9,10 \\
NGC3603 A1 &WN6ha +  & 6.2          &  116$\pm$31          & 3.77$\pm$?    &  $>$20               &   - & - &8,9,10,11     \\
           &WN6ha    &   -           &   89$\pm$16         &               &                      &   &      &     \\
BAT99-99 (Mk39)&   O3If$^*$/WN6 &   -          &      -    & 92.6$\pm$0.3   & 13$\pm$2              &   1.6 & - & 4 \\
WR20a      & WN6ha +  &  6.1          &  82.7$\pm$5.5      &   3.7         &  8.0         & 0.35+1.55  &   -   & 12,13,14 \\
           &  WN6ha   &  6.1          &  81.9$\pm$5.5      &               &                      &       &       &       \\ 
BAT99-118 (R144) & WN6h     &   -          &      -        &   RV var.      & 3.3                   &  2.1  & - & 3,4 \\
WR21a      & WN6ha +  &   -           &  {\em 87$\pm$6}    &  31.7         &  0.4-1.6          & 3.3        &   -   & 15,16   \\
           &  O4     &    -           &  {\em 53$\pm$4}    &               &                      &       &       &       \\
BAT99-119 (R145) & WN6h        &   -          &      -     & 158.8          & 1.0                   &   1.6 & - & 3,17 \\ 
WR22 (HD 92740)       & WN7h  +  &               &  55.3$\pm$7.3      &  80.3         &  0.9                 &   -   &    -  &  18,19 \\
BAT99-103 (R140b)& WN6h         &   -          &      -    & 2.76           & 0.8$\pm$0.3           &   -   & - & 4 \\
WR 148 (HD 197406)    &WN8h  + ? &  -           &      -              & 4.32          &  0.6$\pm$0.3         & - &  -   &  20,21,22 \\ 
BAT99-77   &  WN7ha   &   -          &      -              & 3.00           & 0.5                   &   -   & - & 4 \\
WR12       & WN8h + ?&    -           &   -                &  23.9         &   0.3               &   -   &   -   & 20,23,24\\
BAT99-12   & O3If$^*$/WN6& -          &      -              & 3.23           & $<$6.0                &   -   & - & 4 \\   
BAT99-32   & WN6(h)   &   -          &      -              & 1.91           & $<$4.5                &   -   & - & 4 \\
BAT99-113 (Mk30) & O3If$^*$/WN6  &   -          &      -   &  4.70          & $<$1.3                &   -   & - & 4 \\
BAT99-95 (R135)&   WN7h &   -          &      -            & 2.11           & $<$0.6                &   -   & - & 4 \\
\hline
Arches-F6  & WN9h  &  6.3         &      -             &    -          &   11.0                  &  1.9    &  Y &  25,26 \\
R136a3(bl) & WN5h    &    -          &      -             &    -          &    8.5                  &  4.2    &  - & 3,7 \\
WR20b      & WN6ha   &  5.9          &      -             &    -          &    6.5                  & 0.5+5.5 &  - & 12,13     \\
Arches-F7  & WN9h  &  6.25          &      -             &    -          &     7.2                 &  2.1    &  Y & 25,26 \\
Arches-F9  & WN9h  &  6.3         &      -             &    -          &    4.6                  &  3.3    &  Y & 25,26 \\
\hline
\end{tabular}
\end{center}
{Top panel: summary of the properties of CXO J1745-28 and  X-ray bright ($>10^{35}$ergs$^{-1}$)  binary systems.   
Middle panel: remaining spectroscopically or photometrically confirmed binaries. Bottom panel: binary candidates selected
on the basis of their X-ray  properties (see Sect. 4). Stars are ordered in terms of decreasing X-ray flux, with errors given
 for this parameter and  orbital period if provided in the original study. We quote dynamical masses only, with
those given in italics representing lower limits. Regarding individual sources,
WR20a, 22 and 25 have all been flagged as X-ray variable 
(Naze et al. \cite{naze}, Gosset et al. \cite{gosset} and Pollock \& Corcoran \cite{pollock06}), 
while the X-ray fluxes given for the Arches sources assume a 1T fit; a similar fit for CXO J1745-28 
 results in  a directly comparable flux (Wang et al. \cite{wang}, Mikles et al. \cite{m06}). The converse is also true;  the   
2T (0.9+5.8keV) fit for F9 favoured by Yusef-Zadeh et al. (\cite{yz02}) yields L$_x \sim$ 8$\times$10$^{34}$ergs$^{-1}$; 
we choose to present the 1T fits for all three stars  since Yusef-Zadeh et al. (\cite{yz02}) only present a spectral fit for F9.References for the values quoted in the table are:
(1) Mikles et al. (\cite{m06});
(2) Mikles et al. (\cite{m08});
(3) Townsley et al. (\cite{townsley}); 
(4) Schnurr et al. (\cite{sch08a});
(5) Gamen et al. (\cite{gamen});
(6) Raassen et al. (\cite{raassen});
(7) Schnurr et al. (\cite{sch09b});
(8) Crowther \& Dessart (\cite{crowther});
(9) Schnurr et al. (\cite{sch08});
(10) Moffat et al. (\cite{moffat02});
(11) Moffat et al. (\cite{moffat04});
(12) Naze et al. (\cite{naze});
(13) Tsujimoto et al. (\cite{tsu});
(14) Rauw et al. (\cite{rauw05});
(15) Benaglia et al. (\cite{benaglia});
(16) Niemela  et al. (\cite{niemela});
(17) Schnurr et al. (\cite{sch08});
(18) Schweickhardt et al. (\cite{sch});
(19) Pollock (\cite{pollockE});
(20) Hamann et al. (\cite{hamann}); 
(21) Pollock et al. (\cite{pollockR});
(22) Drissen et al.(\cite{drissen86});
(23) Lamontagne et al. (\cite{lam});
(24) Ignace et al. (\cite{ignace});  
(25) Martins et al. (\cite{martins08}) and 
(26) Wang et al.    (\cite{wang}).}
\end{table*}

\begin{table}
\begin{center}
\caption[]{Summary of confirmed and candidate WNLh binaries  in young open clusters.
The cluster list 
follows that of Crowther et al. (\cite{PAC}), supplemented with the WNLh rich Arches (Figer et al. \cite{figer2002})
and 30 Dor (Walborn \& Blades \cite{nolan}) clusters.  Column 4 summarises
 the number of stars in each cluster for which a RV survey has been carried out. Confirmed binaries were
identified via photometric or spectroscopic evidence, while candidate binaries were selected if  
L$_{x} \geq 10^{34}$ergs$^{-1}$ and/or $kT >4$keV, noting that such strict criteria 
would exclude known spectroscopic binaries such as WR20a and WR22 (Table 4).
Given the different sensitivities of current  RV and X-ray surveys  (e.g. Pollock \cite{pollockE}, Pollock et al. 
\cite{pollockR}, Ignace et al. \cite{ignace}, Oskinova \cite{oskinova}) 
we present results for individual clusters.}
\begin{tabular}{cccccc}
\hline
\hline
 Cluster  &Age &  Total &RV      &\multicolumn{2}{c}{Binaries} \\
         & (Myr)&       & survey &     Confirmed & Candidates \\
\hline
NGC3603    &1.3$\pm$0.3  & 3  & 3 & 2       &    0 \\
Carina     &1.5$\pm$0.5  & 3  & 3 &2       &    0 \\
HM-1       &2.2$\pm$0.5  & 2  & 0 &-       &    0 \\
NGC6231    &2.7$\pm$0.5  & 1  & 0 &-       &    0 \\
Arches     &2.5$\pm$0.5  & 13 & 0 &-       &    3 \\
Wd 2       &2.6$\pm$0.2  & 2  & 1 &1       &    1 \\
Bochum 7   &2.8$\pm$0.5  & 1  & 1 &1       &    0 \\
R136/30 Dor&$>$2         & 23 & 23& 8       &    1 \\        
\hline
Total      &             & 48 & 31& 14      &    5 \\
\hline
\end{tabular}
\end{center}
\end{table}

\section{Discussion and concluding remarks}

We present a tailored non-LTE analysis of the IR counterpart to the bright X-ray source \object{CXO J1745-28}, finding it to be a highly luminous, massive WN8-9h star, with physical parameters comparable to such stars in 
the Arches cluster. Furthermore,  3  of the 4 Arches members with X-ray detections - F6, 7 \& 9 - also share remarkably similar 
X-ray fluxes and spectra to \object{CXO J1745-28}. While the near-IR spectra of these objects
 are consistent with originating in single stars, their  X-ray properties clearly argue for binarity, with a comparison to
known HMXBs and CWBs favouring the latter interpretation  - indeed the Galactic WN6ha+?  CWB system WR25 is  a near twin of \object{CXO J1745-28}. 
Moreover, it appears difficult to reconcile both the  youth and extreme mass inferred for \object{CXO J1745-28} and the Arches 
sources with the requirement for a SNe to have occurred in order  to yield a relativistic companion - however we note that if
 they {\em are} HMXBs then they will provide a unique insight into the final stages of stellar evolution for the most massive 
stars that appear able to form in the local Universe. 

Nevertheless, under either HMXB or CWB  hypothesis these four systems  add to the growing population of binary WNLh stars in the 
galaxy and LMC, which are summarised in Table 4 and currently consist of an additional  19 confirmed and 2 candidates.
With periods ranging from $\sim$1.9-208d, the most compact binaries may be candidates for the Case M 
evolution described by De Mink et al. (\cite{deM}; in which tidal forces spin up the stars leading to significant rotational 
mixing), while the long period systems have yet to encounter mass transfer, thus having evolved as single stars, but ones for 
which dynamical mass estimates may be obtained.

In Table 5 we summarise the population of  WNLh binaries in young massive clusters. This restricted population, rather than
 the complete census, was chosen since it excludes candidates, such as \object{CXO J1745-28}, which have been identified as 
WNLh stars because of their binary derived observational properties and hence would introduce a selection bias.
Thus, of the 31 WNLh stars for which a long term survey for RV variability has been performed, 14 have been 
identified as binaries resulting in a binary fraction of $\sim$45\%, which we consider a lower limit given the 
lack of sensitivity  of current RV surveys to long period systems (e.g. $>$200d \& $>$40d for 30 Dor and R136 
proper;  Schnurr et al. \cite{sch08a},\cite{sch09b}) and the additional X-ray selected candidates. 

This is consistent with other surveys which find  similarly high percentages for OB and 
Wolf-Rayet stars, albeit for different samples of stars comprising either lower (Clark et al. \cite{clark08}; Ritchie et al. \cite{ritchie}) or a wider range 
of masses (Kobulnicky \& Fryer \cite{kobulnicky}, Sana et al. \cite{sana}, Bosch et
 al. \cite{bosch}). Such a binary fraction potentially presents important constraints on the formation mechanisms 
for very massive stars such as \object{CXO J1745-28} and the Arches population. Not only does (accretion driven)
 radiation 
pressure have to be overcome to yield $\geq$80M$_{\odot}$ stars but within a short period of time either the formative processes 
or another mechanism - such as dynamical interactions during cluster core collapse (S. Goodwin, priv. comm. 2009) - yields 
significant numbers of very massive short period systems. 

Finally, referencing the issues raised in the introduction,  the extreme mass inferred for \object{CXO J1745-28}, as well as the 
discovery of similar stars such as \object{WR102ka} (Barniske et al. \cite{barniske}) reinforces the supposition of e.g. 
Mauerhan  et al. (\cite{mauerhan}) that a diffuse, apparently isolated population of massive stars is found within the central 
$\sim$50pc of the Galactic centre in addition to the well know massive clusters. The origin of such stars is currently unclear; 
 at $\sim$17~pc  the {\em projected} distance of \object{CXO J1745-28} from the Arches - the only known GC cluster young enough 
for it to have formed in -  appears uncomfortably large to explain its location as a result of  dynamical ejection (implying a minimum travel 
time of 1.7Myr for a velocity of  10kms$^{-1}$). Finally, as with other massive star forming regions such as 30 Dor 
(Townsley et al. \cite{townsley}), Wd~1 (Clark et al. \cite{clark08}) and the 
putative complex at the base of the Scutum-Crux Arm (Clark et al. \cite{clark09}) there are currently no unambiguous 
HMXB candidates within the Galactic Centre. Despite the high binary fractions inferred for massive stars within these regions,  
the physical processes leading to the 
production of neutron stars and  black holes (binary mass transfer/common envelope evolution and supernovae) appear 
inimical  to the production and/or retention of X-ray bright HMXBs within their natal clusters/complexes.

\begin{acknowledgements}
JSC acknowledges support from an RCUK fellowship, and thanks Julian Pittard, Ollie Schnurr, Selma de Mink, and Fabrice Martins for
their invaluable help during the preparation of this manuscript and John Hillier for the use of his code. 

\end{acknowledgements}

{}

\begin{thebibliography}{}

\bibitem[2004]{asplund}
Asplund, M., Grevesse, N., Sauval, A. J., Allende Prieto, C., Kiselman, D., 2004, A\&A, 417, 751
\bibitem[2006]{audley}
Audley, M. D., Nagase, F., Mitsuda, K., Angelini, L., Kelley, R. L., 2006, MNRAS,
367, 1147
\bibitem[1997a]{babela}
Babel, J., Montmerle, T., 1997a, ApJ, 485, L29
\bibitem[1997b]{babelb}
Babel, J., Montmerle, T., 1997b, A\&A, 323, 121
\bibitem[2008]{barniske}
Barniske, A., Oskinova, L. M., Hamann, W.-R., 2008, A\&A, 486, 971
\bibitem[2005]{benaglia}
Benaglia, P., Romeroa, G. E., Koribalski, B., Pollock, A. M. T., 2005, A\&A, 440, 743 
\bibitem[1999]{bc99}
Bohannan, B., Crowther, P. A., 1999, ApJ, 511, 374
\bibitem[1999]{BAT}
Bosch, G., Terlevich, R., Melnick, J., Selman, F., 1999, A\&AS, 137, 21
\bibitem[2009]{bosch}
Bosch, G., Terlevich, E., Terlevich, R., 2009, AJ, 137, 3437
\bibitem[2008]{bouret}
Bouret, J.-C., Donati, J.-F., Martins, F., et al., 2008, MNRAS, 389 75
\bibitem[2007]{casali07}
Casali, M., Adamson, A., Alves de Oliviera, C., 2007, A\&A, 467, 777
\bibitem[2002]{c02}
Clark, J. S., Goodwin, S. P., Crowther, P. A., et al., 2002,
A\&A, 392, 909
\bibitem[2008]{clark08}
Clark, J. S., Muno, M. P., Negueruela, I., et al., 2008, A\&A, 477, 147 
\bibitem[2009]{clark09}
Clark, J. S., Negueruela, I., Davies, B., 2009, A\&A, 498, 109 
\bibitem[1996]{cotera}
Cotera, A. S., Erickson, E. F., Colgan, S. W. J., et al., 1996, ApJ, 461, 750
\bibitem[2000]{cox}
Cox, A., 2000, Allen's Astrophysical Quantities
\bibitem[1998]{crowther}
Crowther, P. A., Dessart, L., 1998, MNRAS, 296, 622
\bibitem[2006]{PAC}
Crowther, P. A., Hadfield, L. J., Clark, J. S., Negueruela, I., Vacca, W., D.,
2006, MNRAS, 372, 1407
\bibitem[2002]{cui}
Cui, W., Feng, Y. X., Zhang, S. N., et al., 2002, ApJ, 576, 357
\bibitem[2006]{deB}
DeBecker, M., Rauw, G., Sana, H., et al., 2006, MNRAS, 371, 1280
\bibitem[2009]{deM}
De Mink, S.E., Cantiello, M., Langer, N., et al., 2009, A\&A, 497, 243
\bibitem[2002]{donati02}
Donati, J.-F., Babel, J., Harries, T. J., et al., 2002, MNRAS, 333, 55
\bibitem[2006]{donati06}
Donati, J.-F., Howarth, I. D., Bouret, J.-C., et al., 2006, MNRAS, 365, L6
\bibitem[1986]{drissen86}
Drissen, L., Lamontagne, R., Moffat, A. F. J., Bastien, P., Seguin, M., 1986, ApJ, 
304, 188
\bibitem[1995]{figer95}
Figer, D. F., Morris, M., McLean, I. S., 1996, ASPC, 102, 263
\bibitem[2002]{figer2002}
Figer, D. F., Najarro, F.,Gilmore, D., 2002, ApJ, 581, 258 
\bibitem[2005]{gagne}
Gagn\'e, M., Oksala, M. E., Cohen, D. H., et al., 2005, ApJ, 628, 986 
\bibitem[2006]{gamen}
Gamen, R., Gosset, E., Morrell, N et al., 2006, A\&A., 460, 777
\bibitem[2003]{gosset}
Gosset, E., Rauw, G., Vreux, J.-M., et al. 2003, IUAS, 212, 188
\bibitem[2006]{hamann}
Hamann, W. R., Grafener, G., Liermann, A., 2006, A\&A, 457, 1015
\bibitem[2006]{hewett08}
Hewett, P. C., Warren, S. J., Leggett, S. K., Hodgkin, S. T., 2006, MNRAS, 367,
454
\bibitem[1998]{hil98}
 Hillier, D.~J.~\& Miller, D.~L.\ 1998, ApJ, 496, 407
\bibitem[1999]{hil99} 
Hillier, D.~J.~\& Miller, D.~L.\ 1999, ApJ, 519, 354
\bibitem[2003]{h03}
Hillier, D. J., Lanz, T., Heap, S. R., et al., 2003, ApJ, 588, 1039 
\bibitem[2009]{hodgkin09}
Hodgkin, S., T., Irwin, M. J., Hewett, P. C., Warren, S. J., 2009, MNRAS, 394, 675
\bibitem[2008]{hubrig}
Hubrig, S., Scholler, M., Schnerr, R. S., et al., 2008, A\&A, 490, 793
\bibitem[2000]{ignace}
Ignace, R., Oskinova, L. M., Foullon, C., 2000, MNRAS, 318, 214
\bibitem[2005]{i05}
Indebetouw, R., Mathis, J. S., Babler, B. L., 2005, ApJ, 619, 931
\bibitem[2006]{kaper}
Kaper, L., van der Meer, A., Najarro, F., 2006, A\&A, 457, 505
\bibitem[2006]{kim06}
Kim, S. S., Figer, D. F., Kudritzki, R. P., Najarro, F., 2006, ApJ, 653 L113
\bibitem[2007]{kobulnicky}
Kobulnicky, H. A., Fryer, C. L., 2007, ApJ, 670, 747
\bibitem[1995]{krabbe}
Krabbe, A., Genzel, R., Eckart, A., et al., 1995, APJ, 447, L95
\bibitem[2005]{lab}
La Barbera, A., Segreto, A., Santangelo, A., Kreykenbohm I., Orlandini, M.,
2005, A\&A, 438, 617
\bibitem[1996]{lam}
Lamontagne, Moffat, A. F. J., Drissen, L., et al., 1996, AJ, 112, 2227 
\bibitem[2003]{lh03}
Lanz, T., Hubeny, I., 2003, ApJS, 146, 417
\bibitem[2007]{ukidss}
Lawrence, A., Warren, S. J., Almaini, O., et al., 2007, MNRAS, 379, 1599
\bibitem[2005]{martins05}
Martins, F., Schaerer, D., Hillier, D. J., 2005, A\&A, 441, 735  
\bibitem[2006]{mp06}
Martins, F., Plez, B., 2006, A\&A, 457, 637 
\bibitem[2008]{martins08}
Martins,F., Hillier, D. J., Paumard, T., 2008, A\&A, 478, 219
\bibitem[2009]{mason}
Mason, A. B., Clark, J. S., Norton, A. J., Negueruela, I., Roche, P., 2009, A\&A, 
in press, astro-ph/0907.3876
\bibitem[2007]{mauerhan}
Mauerhan, J. C., Muno, M. P., Morris, M., 2007, ApJ, 662, 574
\bibitem[2000]{meynet}
Meynet, G., Maeder, A., 2000, A\&A, 361, 101
\bibitem[2006]{m06}
Mikles, V. J., Eikenberry, S. S.,Muno, M. P.,  Bandyopadhyay, R. M., Patel, S.,
2006, ApJ, 651, 408
\bibitem[2008]{m08}
Mikles, V. J., Eikenberry, S. S., Bandyopadhyay, R. M., Muno, M. P., 
2008, ApJ, 689, 1222
\bibitem[2002]{moffat02}
Moffat, A. F. J., Corcoran, M. F., Stevens, I. R., et al., 2002, ApJ, 573, 191 
\bibitem[2004]{moffat04}
Moffat, A. F. J., Poitras, V., Marchenko, S. V.,   et al., 2004, AJ, 128, 2854
\bibitem[2004]{moffat08}
Moffat, A. F. J., 2008, RMxAC, 33, 95
\bibitem[2008]{montes}
Montes, G., Perez-Torres, M. A., Alberdi, A., Gonzalez, R. F., 2008, arXiv0810.5026
\bibitem[2000]{morris}
Morris, P. W., van der Hucht, K. A., Crowther, P. A., 2000, A\&A, 353, 624
\bibitem[2006a]{muno06a}
Muno, M. P., Bauer, F. E., Bandyopadhyay, R. M., Wang, Q. D., 2006a,
ApJS, 165, 173
\bibitem[2006b]{muno06b}
Muno, M. P., Bower, G. C., Burgasser, A. J., et al., 2006b, ApJ, 638, 183 
\bibitem[2009]{muno09}
Muno, M. P., Bauer, F. E., Baganoff, F. K., 2009, ApJS, 181, 110
\bibitem[1995]{nagata}
Nagata, T., Woodward, C. E., Shure, M., Kobayashi, N., 1995, AJ, 109, 1676
\bibitem[2004]{naze04}
N\'aze, Y., Rauw, G., Vreux, J.-M., DeBecker, M., 2004, A\&A, 417, 667
\bibitem[2007]{naze07}
N\'aze, Y., Rauw, G., Pollock, A. M. T., Walborn N. R., Howarth, I. D., 2007, MNRAS, 375, 145
\bibitem[2008a]{naze}
N\'aze, Y., Rauw, G., Manfroid, J., 2008b, A\&A, 483, 171
\bibitem[2008b]{naze08}
N\'aze, Y., Walborn, N. R., Martins, F., 2008b, RMxAA, 44, 331
\bibitem[2008c]{naze08b}
N\'aze, Y., Walborn, N. R., Rauw, G. et al., 2008c, AJ, 135, 1946
\bibitem[2008]{niemela}
Niemela, V. S., Gamen, R. C., Barba, R. H., et al., 2008, MNRAS, 389, 1447
\bibitem[2005]{oskinova}
Oskinova, L. M., 2005, MNRAS, 361, 679
\bibitem[2002]{pittards}
Pittard, J. M., Stevens. I. R., 2002, A\&A, 388, L20
\bibitem[2002]{pittardc}
Pittard, J. M., Corcoran, M. F., 2002, A\&A, 383, 636
\bibitem[1987]{pollockE}
Pollock, A. M. T., ApJ, 320, 283
\bibitem[1995]{pollockR}
Pollock, A. M. T., Haberl, F., Corcoran, M. F., 1995, IAUS, 163, 522 
\bibitem[2006]{pollock06}
Pollock, A. M. T., Corcoran, M. F., 2006, A\&A, 1093
\bibitem[1990]{prinja}
Prinja, R. K., Barlow, M. J., Howarth, I. D., 1990, ApJ, 361, 607
\bibitem[2002]{pz}
Portegies Zwart, S. F., Pooley, D., Lewin, W. H. G.,2002, ApL, 574, 762
\bibitem[2003]{raassen}
Raassen, A. J. J., van der Hucht, K. A., Mewe, R., 2003, A\&A, 402, 653
\bibitem[2008]{raassenmag}
Raassen, A. J. J., van der Hucht, K. A.,  Miller, N. A., Cassinelli, J. P., 2008, A\&A, 478, 513 
\bibitem[1996]{rauw96}
Rauw, G., Vreux, J.-M., Gosset, E., et al., 1996, A\&A, 306, 771
\bibitem[2002]{rauw02}
Rauw, G., Blomme, R., Walborn, W. L., et al., 2002, A\&A, 394, 993
\bibitem[2005]{rauw05}
Rauw, G., Crowther, P. A., De Becker, M., et al., 2005, A\&A, 432, 985
\bibitem[2003]{r03}
Rayner, J. T., Toomey, D. W., Onaka, P. M., et al., 2003, PASP, 115, 362
\bibitem[1993]{1993}
Reid, M. J., 1993, ARA\&A, 31, 345
\bibitem[2009]{ritchie}
Ritchie, B. W., Clark, J. S., Negueruela, I., Crowther, P. A., 2009, A\&A, in press, arXiv:0909.3815v1 [astro-ph.SR]
\bibitem[2008]{sana}
Sana, H., Gosset, E., Naze, Y., Rauw, G, Linder, N., 20008, MNRAS, 386, 447
\bibitem[2008a]{sch08a}
Schnurr, O., Moffat, A. F. J., Villar-Sbaffi, A., St-Louis, N., Morrell, N. I., 2009, MNRAS,
395, 823 
\bibitem[2008b]{sch08}
Schnurr, O., Casoli, J., Chene, A.-N., Moffat, A. F. J., St-Louis, N., 
2008b, MNRAS, 389, L38
\bibitem[2009a]{sch09}
Schnurr, O., Moffat, A. F. J., Villar-Sbaffi, St-Louis, N., Morrell, N. I., 
2009a, MNRAS, 395, 823
\bibitem[2009b]{sch09b}
Schnurr, O., Chene, A.-N., Casoli, J., Moffat, A. F. J., St-Louis, N., 2009, MNRAS, submitted
\bibitem[1999]{sch}
Schweickhardt, J., Schmutz, W., Stahl, O., Szeifert, Th, Wolf, B., 1999, A\&A, 347, 127
\bibitem[1992]{stevens}
Stevens, I. R., Blondon, J. M., Pollock, A. M. T., 1992, ApJ, 386, 265
\bibitem[2006]{townsley}
Townsley, L. K., Broos, P. S., Feigelson, E. D., Garmire, G. P., Getman, K. V.,
2006, AJ, 131, 2164
\bibitem[2007]{tsu}
Tsujimoto, M., Feigelson, E. D., Townsley, L. K., 2007, ApJ, 665, 719
\bibitem[2002]{yz02}
Yusef-Zadeh, F., Law C., Wardle, M., et al., 2002, ApJ, 570, 665
\bibitem[1982]{w82}
Walborn, N. R., 1982, ApJ, 256, 452
\bibitem[1989]{nolan}
Walborn, N. R., Blades, J. C., 1997, ApJS, 112, 457
\bibitem[2006]{wang}
Wang, Q., D., Dong, H., Lang, C., 2006, MNRAS, 371, 38
\bibitem[1983]{white}
White, N. E., Kallman, T. R., Swank, J. H., 1983, ApJ, 269, 264
\end{thebibliography}
\end{document}